# Imaging Magnetic Focusing of Coherent Electron Waves

Katherine E. Aidala,[1*] Robert E. Parrott,[2] Tobias Kramer,[2] E.J. Heller,[2,3]

R.M. Westervelt,[1,2] M.P. Hanson[4] and A.C. Gossard[4]

[1]*Div. of Engineering and Applied Sciences, Harvard University, Cambridge, MA 02138*

[2]*Dept. of Physics, Harvard University, Cambridge, MA 02138*

[3]*Dept. of Chemistry and Chemical Biology, Harvard University, Cambridge, MA 02138*

[4]*Dept. of Materials Science, University of California, Santa Barbara, CA 93106*

[*] Present Address: Dept. of Physics, Mount Holyoke College, South Hadley, MA 01075

**The magnetic focusing of electrons has proven its utility in fundamental studies of electron transport.[1-4] Here we report the direct imaging of magnetic focusing of electron waves, specifically in a two-dimensional electron gas (2DEG). We see the semicircular trajectories of electrons as they bounce along a boundary in the 2DEG, as well as fringes showing the coherent nature of the electron waves. Imaging flow in open systems is made possible by a cooled scanning probe microscope.[5-17] Remarkable agreement between experiment and theory demonstrates our ability to see these trajectories and to use this system as an interferometer. We image branched electron flow[11] as well as the interference of electron waves.[10,11,18] This technique can visualize the motion of electron waves between two points in an open system, providing a straightforward way to study systems that may be useful for quantum information processing and spintronics.[19-21]**

For nanoscale devices at low temperatures, both the particle and wave aspects of electron motion are important. Electrons can travel ballistically through an unconfined two-dimensional electron gas (2DEG) while their quantum phase remains coherent. The development of unconfined or open devices in this regime[1,22-26] clears the way for new applications in electron interferometry, spintronics, and quantum information processing.[18-21,27] The most basic question for open systems is: "What is the pattern of flow for electron waves?" Imaging flow is difficult, because the electrons are often buried inside a heterostructure, and because low temperatures and strong magnetic fields are needed to observe quantum phenomena.

A cooled scanning probe microscope (SPM) provides a way to image flow: a conducting SPM tip above the sample capacitively couples to the 2DEG below.[5-17] Pathways for electron waves emerging from a quantum point contact (QPC) were revealed[10] and found to have a dramatic branched form.[11] The branching was shown to be the real-space consequence of small-angle scattering by charged donor atoms - the well known mechanism that limits the electron mobility. A charged tip was able to image this flow, because it backscattered electrons; some of them follow a time-reversed path back through the QPC, measurably reducing its conductance. Images of flow were obtained by displaying the QPC conductance as the tip was scanned in a plane above the sample.[10,11] However, this backscattering imaging technique cannot be used in a perpendicular magnetic field, because the electrons follow curved paths and no longer reverse. Magnetic fields are important for spintronics and quantum information processing,[19-21] so it is important to understand electron flow with a field present.

In this letter, we demonstrate a new way to image the flow of electron waves between two points in an open system using a cooled SPM, and we use this approach to image magnetic focusing of electron waves in a 2DEG. Magnetic focusing plays an important role in the study of ballistic transport.[1-4] Images were obtained by using an electron lens formed in the 2DEG beneath a charged SPM tip to redirect flow, by throwing a shadow downstream. We obtain clear pictures of semicircular bouncing-ball patterns of particle flow that cause magnetic focusing. Branching of the flow is visible. In addition, we see fringes created by the interference of electron waves bouncing off the tip with those travelling directly between the two QPCs. In this way, the cooled SPM acts as a new type of electron interferometer for solids. The observed bouncing-ball patterns of flow and interference fringes are in excellent agreement with full quantum simulations that include the tip-induced electron lens and small-angle scattering, demonstrating that this new imaging technique accurately views the flow of electron waves through the device.

Figure 1 illustrates the device geometry and the imaging technique. Two QPCs are formed in a 2DEG by surface gates (Fig. 1a); a scanning electron micrograph of the device is shown in Fig. 1b. The separation between the QPC centers is $L = 2.7$ µm. A 2DEG with density $3.8 \times 10^{11}$ cm$^{-2}$ and mobility 500,000 cm$^2$/Vs is located 47 nm beneath the surface of a GaAs/AlGaAs heterostructure with the following layers: 5 nm GaAs cap, 20 nm Al$_{0.3}$Ga$_{0.7}$As barrier, Si delta-doping, then a 22 nm Al$_{0.3}$Ga$_{0.7}$As barrier next to the 2DEG in GaAs. Metal gates that define the QPCs were fabricated using e-beam lithography. The device is mounted in the SPM, inside a superconducting solenoid, and cooled to 4.2 K. A computer controls the SPM and records the images.

Magnetic focusing (Fig. 1d) occurs because electrons leaving one QPC over a range of angles circle around and rejoin at the second QPC, when the spacing $L$ is close to the diameter of a cyclotron orbit. For GaAs the electron cyclotron orbit is circular with radius:

$$r_c = \hbar k_F / eB,  \tag{1}$$

where $e$ is the electron charge, $B$ is the perpendicular magnetic field, and $k_F$ is the Fermi wavevector. As $B$ is increased, the first focusing peak occurs when $L = 2r_c$. Additional peaks occur at higher fields when $L = 2nr_c$ is an integer multiple of the cyclotron diameter at fields:

$$B_n = 2n\hbar k_F / eL.  \tag{2}$$

The shape, clarity and spacing of magnetic focusing peaks provide information about ballistic flow in the sample material.  The effects of small-angle scattering are shown in the simulation of Fig. 1e (and in supplementary material):  magnetic focusing still occurs, but the flow now contains branches[11] similar to flow in the SPM image in Fig. 1b for $B = 0$.

To image electron flow from one point to another, a small movable electron lens is created by creating a dip in electron density immediately below the charged SPM tip.  The lens deflects electrons (Fig. 1c), throwing a V-shaped shadow downstream.  An image of electron flow is obtained by displaying the transmission $T$ between the QPCs as the tip is scanned across a plane 10 nm above the surface.  $T$ is measured by recording the voltage across the second QPC as a known current is passed through the first.  When electrons are ballistically injected into the second QPC from the first, a voltage develops that drives a current in the opposite direction to cancel the influx of electrons; this voltage is proportional to $T$.

Simulations of electron flow that show how the imaging technique works are presented in Figs. 1c-f.  A Gaussian $\varphi_{tip} = V_o \exp\left(-\left(r - r_{tip}\right)^2 \big/ 2a^2\right)$ is used to model the tip potential in the 2DEG, where $r_{tip}$ is the tip position, $a$ is the width, and $V_o$ is the height.  For this paper $V_o > 0$, and the tip creates a dip in electron density.  The relative strength of the tip potential is

$$\eta = \frac{V_o}{E_F},$$   (3)

where $E_F$ is the Fermi energy.  For $\eta < 1$, the electron gas is partially depleted, and an imperfect diverging lens is created with a focal length determined by $\eta$.  When $\eta \geq 1$, the 2DEG is fully depleted, and electrons can backscatter.  Simulations (Fig. 1c) at $B = 0$ show how a weak lens ( $\eta = 0.2$ ) creates a V-shaped shadow downstream by forcing electrons to the sides, where they form two caustics in flow along the legs of the V.  Figure 1f shows how we image magnetic focusing: when the shadow cast by the lens beneath the tip hits the second QPC, $T$ is reduced. The amount of reduction $\Delta T$ is proportional to the original flux, before the tip was introduced. By displaying $\Delta T$ *vs.* tip position $r_{tip}$ as the tip is raster scanned in a plane above, an image of the original flow is created.[28]

Experimental images of magnetic focusing near the first, second, and third peaks are presented in Figs. 2a-c for a weak tip ( $\eta \approx 0.5$ ), and in Figs. 2d-e for a strong tip ( $\eta \approx 1.0$ ). The bouncing semicircular cyclotron orbits imaged here are an experimental visualization of the origins of magnetic focusing. Figure 2a shows a single, semicircular crescent characteristic of the first peak. Branches in flow from small-angle scattering are also visible. For Fig. 2b the bounce characteristic of the second peak is clearly seen. For Fig. 2c, recorded near the third focusing peak, it becomes difficult to see distinct bounces, although circular features are evident, with radii comparable to $r_c$. The bouncing-ball orbits begin to form the semi-classical equivalent of an edge state.

With a strongly scattering tip ( $\eta \sim 1.0$ ) the SPM images (Figs. 2d-f) have new and distinctly finer features, resulting from the interference of electron waves, some traveling along new pathways created by scattering at the tip. In the absence of the tip, electrons move from one QPC to the other by simultaneously traveling along multiple paths, which add up with a particular overall phase. As quantified below, a strongly scattering tip introduces new trajectories (and removes some of the old). These new trajectories interfere with the original ones, and create interference fringes as the tip is scanned that can be seen in the experimental images. The images in Figs. 2d-f also show bouncing ball orbits similar to their counterparts in Figs. 2a-c. The striking difference is the appearance of narrow fringe-like features. In some locations, a noticeably periodic structure of fringes exists, shown in the blowups Figs. 2g and 2h.

We can understand both the classical and quantum behavior by using full, thermally averaged quantum simulations of an SPM image including tip scattering. Figures 3a-c show simulations of an image on the first magnetic focusing peak for weak ($\eta = 0.2$), moderate ($\eta = 0.6$) and strong ($\eta = 1.2$) tips. These were obtained at 1.7 K using a thermal wavepacket calculation (26) with $\lambda_F = 40$ nm, and they include a random background potential from donor ions. For a weak tip in Fig. 3a, the dark area of reduced transmission ($\Delta T < 0$) corresponds to a classical cyclotron orbit. For moderate and strong tips in Figs. 3b and 3c, quantum interference fringes created by tip scattering become visible. The increase in fringe-like structure as the tip strength is increased is in excellent qualitative agreement with the behavior of the experimental SPM images in Figs. 2a-c and Figs. 2d-f.

To understand the source of contrast in the experimental images, it is very useful to compare quantum simulations with classical trajectories.[10-11] Figure 3d joins quantum simulations from Fig. 3a (beige surface) with classical trajectories (red lines) computed without the tip present using ray tracing. The background potential is shown in blue. Areas with $\Delta T < 0$ are eliminated for ease of comparison. We see that the original trajectories line up with paths of decreased transmission, because the flow is blocked by the tip. These simulations reveal the unusual, and yet very informative way that electron flow is encoded in the experimental images.

The origin of fringing can be understood by a simple semiclassical argument pictured in Fig. 4a. When a new trajectory with phase $\phi$ deflects from the tip and reaches the target QPC, it contributes an amplitude $ae^{i\phi}$ which interferes coherently with the background amplitude $A_o e^{i\phi_o}$ from the nascent trajectories, such that the change in transmission depends on the phase difference $\Delta T \propto \cos(\phi - \phi_o)$. This is illustrated in Fig. 4a by the interference of a direct path between the QPCs along a single cyclotron orbit, with a path deflected by the tip, composed of two cyclotron orbit segments. The phase of the deflected trajectory $\phi = S/\hbar$ is proportional to the classical action $S$ accumulated along the trajectory. When the tip is moved, $S$ changes, and so does $\phi$. This leads to a simple equation for the fringe spacing $d$:

$$d = \frac{\lambda_F}{2} \csc \frac{\theta}{2} \quad , \tag{4}$$

where $\theta$ is the angle by which the trajectory was deflected by the tip, shown in Fig. 4a. When the tip backscatters by $\theta = \pi$, the fringe spacing is $\lambda_F / 2$ as has been seen in previous experiments.[10,11] When using a weak tip with $\eta < 1$, the maximum possible scattering angle is $\theta < \pi$, because it cannot backscatter. This implies that the minimum fringe spacing is $d_{\min} > \lambda_F / 2$. For sufficiently weak tips $\eta \sim 0.1$, the fringe spacings are so large that they are indistinguishable from classical structures such as branches.

A direct comparison of fringing between theory (Figs. 4b-d) and experiment (Figs. 4e-g) shows remarkable agreement. A series of quantum simulations in a small region (outlined in black in Fig. 3c) is compared with SPM images of a comparable area. The simulations in Figs. 4b-d are for $\eta = 0.2$, 0.6 and 1.0 respectively, with the corresponding experimental images for these tip strengths displayed below in Figs. 4e-g. The agreement is excellent. The fringe spacing for a strong tip is comparable to $\lambda_F$, to be expected when the electrons are scattered by $\theta \sim \pi/3$. In this geometry the SPM acts as an interferometer that could be used to extract information from the fringes about the momentum and energy of the electrons.

We have successfully visualized coherent electron transport patterns in a 2DEG magnetic focusing experiment and provided theoretical explanations of phenomena not previously predicted nor measured. The complex interaction of focusing, branching, and tip scattering has been unraveled, revealing the true nature of electron pathways in a real device.

**References**

1. van Houten, H. *et al.* Coherent electron focusing with quantum point contacts in a two-dimensional electron gas. *Phys. Rev. B* **39**, 8556 (1989).

2. Sharvin, Y.V., & Fisher, L.M. Observation of focused electron beams in a metal. *JETP Letters* **1**, 152 (1965).

3. Tsoi, V. Focusing of electrons in a metal by a transverse magnetic field. *JETP Letters* **19**, 70 (1974).

4. Rokhinson, L.P., Larkina, V., Lyanda-Geller, Y.B., Pfeiffer, L.N., & West, K.W. Spin Separation in Cyclotron Motion. *Phys. Rev. Lett.* **93**, 146601 (2004).

5. Topinka, M.A., Westervelt, R.M., & Heller, E.J. Imaging Electron Flow *Physics Today* **56**, 47 (2003), and references therein.

6. Eriksson, M.A. *et al.* Cryogenic scanning-probe characterization of semiconductor nanostructures. *Appl. Phys. Lett.* **69**, 671-673 (1996).

7. Tessmer, S.H., Glicofridis, P.I., Ashoori. R.C., Levitov, L.S., & Melloch, M.R. Subsurface charge accumulation imaging of a quantum Hall liquid. *Nature* **392**, 51 (1998).

8. Yacoby, A., Hess, H.F., Fulton, T.A., Pfeiffer, L.N., & West, K.W. Electrical imaging of the quantum Hall state. *Solid State Commun.* **111**, 1 (1999).

9. McCormick, K.L. *et al.* Scanned potential microscopy of edge and bulk currents in the quantum Hall regime. *Phys. Rev. B* **59**, 4654-4657 (1999).

10. Topinka, M.A., *et al.* Imaging Coherent Flow from a Quantum Point Contact. *Science* **289**, 2323-2326 (2000).

11. Topinka, M.A., *et al.* Coherent Branched Flow in a Two-Dimensional Electron Gas. *Nature* **410**, 183 (2001).

12. Steele, G.A., Ashoori, R.C., Pfeiffer, L.N, & West, K.W. Imaging transport resonances in the quantum Hall effect. *Phys. Rev. Lett.* **95** 136804 (2005).

13. Crook, R., Smith, C.G., Simmons, M.Y., & Ritchie, D.A. Imaging cyclotron orbits and scattering sites in a high-mobility two-dimensional electron gas. *Phys. Rev. B* **62**, 5174-5178 (2000).

14. Crook, R., *et al.* Erasable electrostatic lithography for quantum computers. *Nature* **424**, 751, (2003).

15. Ihn, T., *et al.* Local spectroscopy of edge channels in the quantum Hall regime with local probe techniques. *Physica E* **13**, 671-674 (2002).

16. Kicin, S., *et al.* Local backscattering in the quantum Hall regime. *Phys Rev. B* **70** 205302 (2004).

17. Aoki, N., da Cunha, C.R., Akis, R., Ferry, D.K., & Ochiai, Y. Imaging of integer quantum Hall edge state in a quantum point contact via scanning gate microscopy. *Phys. Rev. B* **72** 155327 (2005).

18. LeRoy, B.J., *et al.* Imaging Electron Interferometer. *Phys. Rev. Lett.* **94**, 126801 (2005).

19. Awschalom, D.D., Loss, D., & Samarth, N. *Semiconductor Spintronics and Quantum Computation.* Springer (2002).

20. Saraga, D.S., Altshuler, B.L., Westervelt, R.M., & Loss, D. Coulomb Scattering in a 2D Interacting Electron Gas and Production of EPR Pairs. *Phys. Rev. Lett.* **92**, 246803 (2004).

21. Schliemann, J., Loss, D., & Westervelt, R.M. *Zitterbewegung* of Electronic Wave Packets in III-V Zinc-Blende Semiconductor Quantum Wells. *Phys. Rev. Lett.* **94**, 206801 (2005).

22. Sohn, L.L., Kouwenhoven, L.P., & Schon, G. *Mesoscopic Electron Transport.* Kluwer Academic Publishers (1997), and references therein.

23. van Wees, B.J., *et al.* Quantized conductance of point contacts in a two-dimensional electron gas. *Phys. Rev. Lett.* **60**, 848 (1988).

24. Shepard, K.L., Roukes, M.L., & Van der Gaag, B.P. Direct measurement of the transmission matrix of a mesoscopic conductor. *Phys. Rev. Lett.* **68**, 2660 (1992).

25. Crommie, M.F., Lutz, & C.P., Eigler, D.M. Confinement of Electrons to Quantum Corrals on a Metal Surface. *Science* **262**, 218 (1993).

26. Kikkawa, J.M., & Awschalom, D.D. Lateral drag of spin coherence in gallium arsenide. *Nature* **397**, 139 (1999).

27. Katine, J.A., *et al.,* Point Contact Conductance of an Open Resonator. *Phys. Rev. Lett.* **79**, 4806 (1997).

28. Aidala, K.E., Parrott, R.E., Heller, E.J., & Westervelt, R.M. Imaging electrons in a magnetic field. *Physica E* **34**, 409 (2006).

29. Heller, E.J. *et al.* Thermal averages in a quantum point contact with a single coherent wave packet. *Nano Letters* **5**, 1285 (2005).

**Author Contributions:** Katherine E. Aidala conducted the experiments with R.M. Westervelt; Robert E. Parrot and Tobias Kramer did classical and quantum simulations of electron flow with E.J. Heller, and M.P. Hanson grew the semiconductor heterostructure with A.C. Gossard.

**Acknowledgements:** This work has been performed with support at Harvard University from the ARO, the NSF-funded Nanoscale Science and Engineering Center (NSEC), and the DFG (Emmy-Noether program). Work at Santa Barbara has been supported in part by the Institute for Quantum Engineering, Science and Technology (iQUEST). We would also like to thank the National Nanotechnology Infrastructure Network (NNIN) and the Harvard CrimsonGrid for computing resources.

**Correspondence** and requests for materials should be addressed to R.M.W. (e-mail: westervelt@deas.harvard.edu).

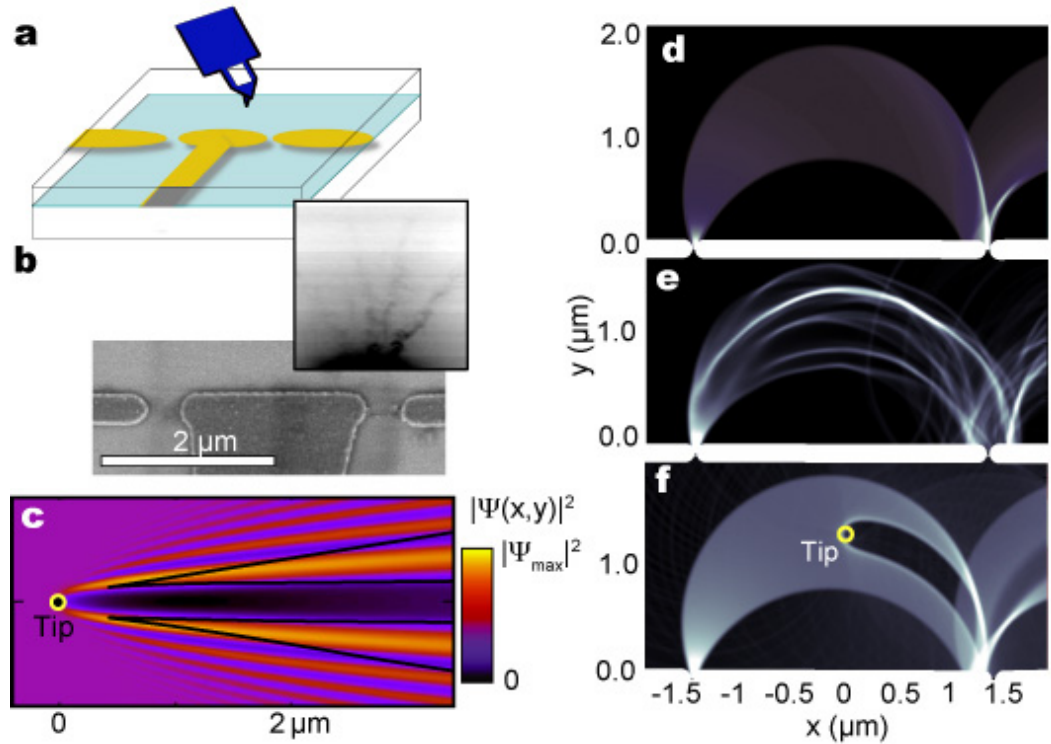

Figure 1 (a) Schematic showing a conducting scanning probe microscope (SPM) tip scanned above the surface of a device. (b) Scanning electron micrograph of the magnetic focusing device; the inset shows an SPM image of the electron flow at zero magnetic field, displaying branches in electron flow. (c) Quantum simulation showing the scattering of an incoming plane wave by an SPM tip, with 2DEG Fermi energy $E_F$ = 13 meV, and a Gaussian tip potential with height 0.2 $E_F$ and width 50 nm. (d-f) Classical simulations of magnetic focusing in a magnetic field: (d) in a flat potential, showing circular cyclotron orbits, (e) with small-angle scattering, showing added branches, and (f) in a flat potential with an SPM tip.

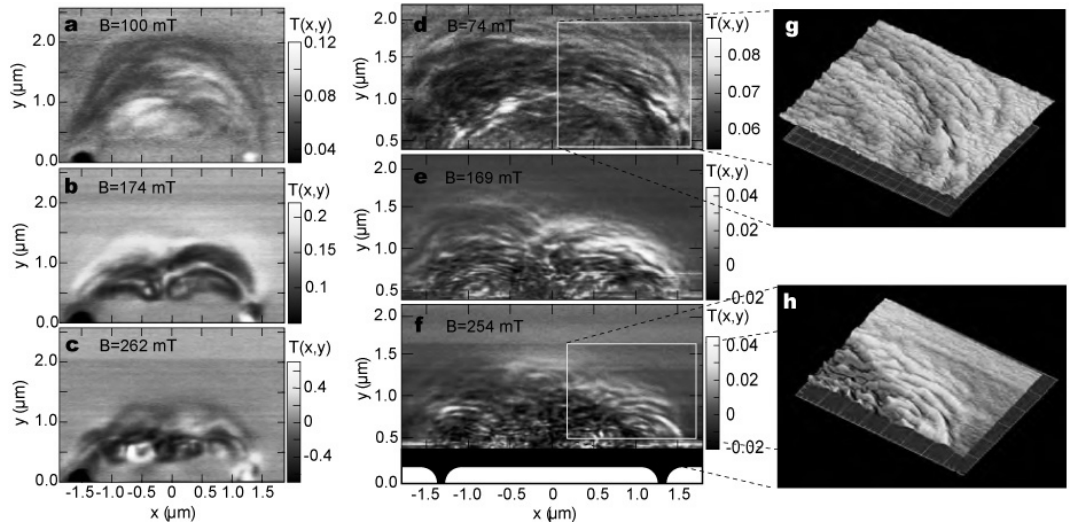

Figure 2 Experimental SPM images of magnetic focusing in a 2DEG at 4.2 K recorded near the first three magnetic focusing peaks: (a-c) Weakly focusing ($\eta \approx 0.5$) tip 90 nm above the 2DEG recorded at $B$ = 100 mT, $B$ = 174 mT, and $B$ = 262 mT with cyclotron radii $r_c$ = 970 nm, $r_c$ = 560 nm and $r_c$ = 370 nm, respectively. The left QPC is on the first conductance plateau, the right is on the third. (d-f) Strongly focusing ($\eta \approx 1.0$) tip 60 nm above the 2DEG recorded at $B$ = 74 mT, $B$ = 169 mT, and $B$ = 254 mT with cyclotron radii $r_c$ = 1310 nm, $r_c$ = 580 nm and $r_c$ = 380 nm, respectively. Both QPCs are on the second conductance plateau. The color scale shows the change in transmission between the QPCs induced by the tip. (g,h) Closeup surface plots in the yellow rectangles of (d,f) that show the regularity and consistency of the quantum fringe structure.

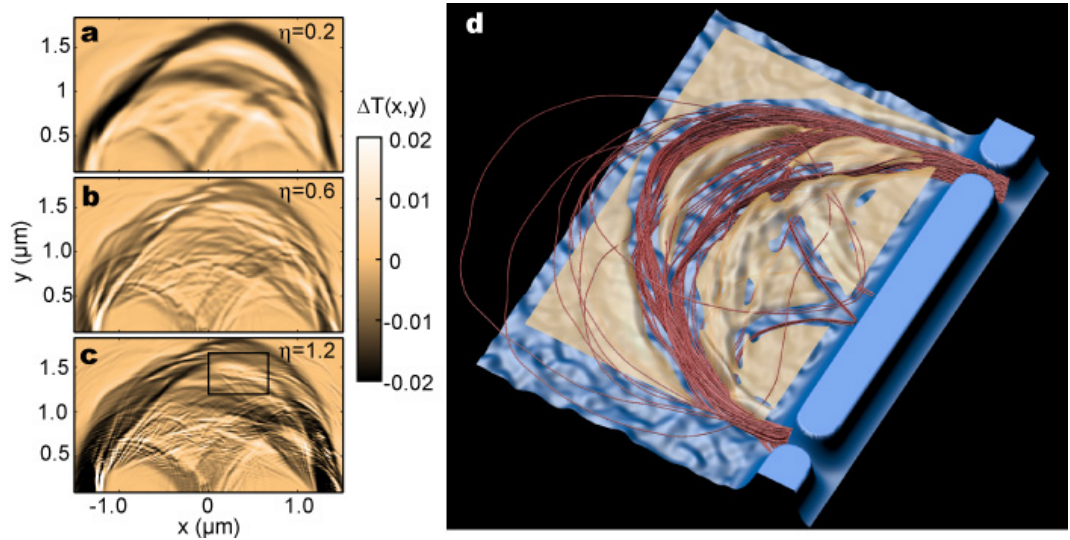

Figure 3  Quantum simulations of SPM images of the first magnetic focusing peak ($B$ = 77 mT) for $\lambda_F$ = 40 nm at 1.7 K, showing the change in transmission $\Delta T$ between QPCs as the tip is scanned above, including small-angle scattering.  (a) For a weak tip ($\eta$ = 0.2) that scatters into small angles, the dark area of reduced transmission ($\Delta T < 0$) shows a classical cyclotron orbit.  For (b) moderate ($\eta$ = 0.6) and (c) strong ($\eta$ = 1.2) tips, quantum interference fringes become visible.  (d) Correspondence between the simulated SPM image (beige surface) from (a) and ray tracing calculations of the originally transmitted electron trajectories (red lines) before the tip was present; regions with $\Delta T < 0$ are omitted for ease of comparison. The blue surface is the smoothly disordered background potential.  The dark areas with reduced transmission ($\Delta T < 0$) line up very well with the original classical trajectories. For stronger tips, imaging of the original classical trajectories becomes more difficult, but fringing reveals regions of high coherent flux of electron waves.

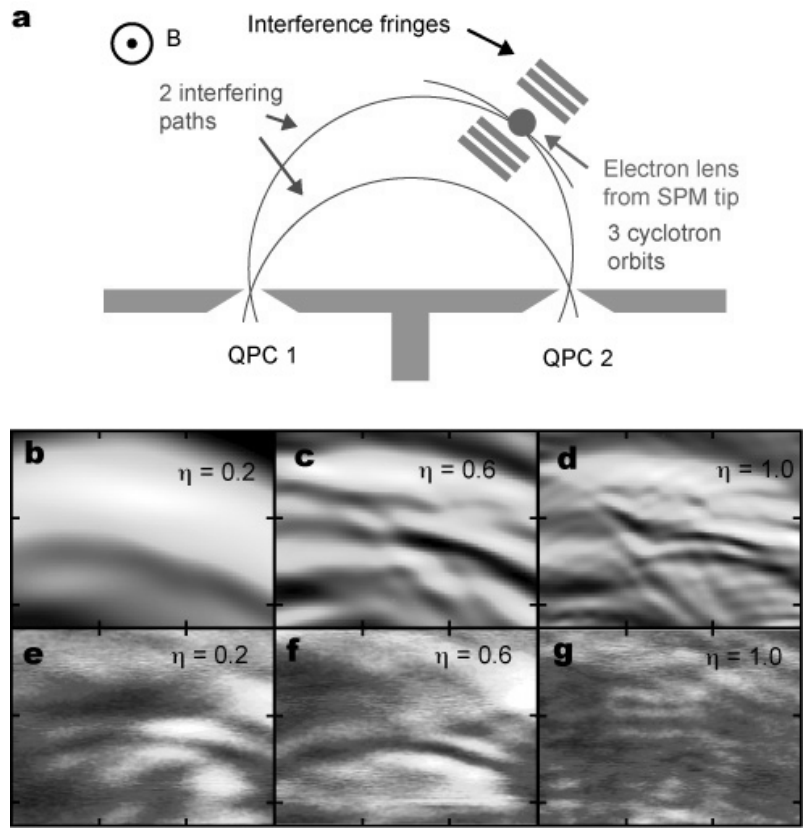

Figure 4   Direct comparison of interference fringes between experiment and theory for different tip strengths. (a) How interference occurs between a path deflected by the SPM tip with a direct path between the two QPCs.  (b-d) Quantum simulations of interference fringes at 1.7 K and $B = 77$ mT for weak ($\eta = 0.2$), moderate ($\eta = 0.6$) and strong ($\eta = 1.0$) tip strengths; the Fermi wavelength is $\lambda_F = 40$ nm.  The panels are located in the black box in Fig. 3c.  (e-g) SPM images at $B = 173$ T showing fringes that appear and move closer as the tip strength $\eta$ increases, in good agreement with the simulations.  The images have dimensions 600 x 450 nm$^2$ and are located 750 nm above and 500 nm to the right of the midpoint between the two QPCs.